\documentclass[12pt]{article}
\usepackage{amssymb,amsmath}
\begin{document}

                            \begin{center}
                           
\LARGE\bf{Tsallis entropy and hyperbolicity}\\

                            \vspace{8mm}

              \Large\rm{Nikos \ Kalogeropoulos}\\

                            \vspace{2mm}
                            
 \normalsize\emph{Weill Cornell Medical College in Qatar,
 Education City,  P.O.  Box 24144, Doha, Qatar\\
                 nik2011@qatar-med.cornell.edu}

                            \end{center}

    \normalsize\rm\setlength{\baselineskip}{15pt} 

                                                           \vspace{5mm}

\noindent {\bf Abstract}. \ Some preliminary evidence suggests the conjecture that the collective 
behaviour of systems having long-range interactions may be described more effectively 
by the Tsallis rather than by the Boltzmann/Gibbs/Shannon entropy. To this end, we examine 
consequences of the biggest difference between these two entropies: their 
composition properties. We rely on a metric formalism that establishes the 
``hyperbolic" nature of Tsallis entropy and explore some of its consequences 
for the underlying systems. We present some recent and some forthcoming results of our work.\\ 
 
                                                       \vspace{8mm}        

%                             \vfill
%
%\noindent\sf  MSC:  \  \  \  \  \  \  51P05, \  53Z05, \ 58Z05, \ 82B99.          \\
%\noindent\sf Keywords:  Tsallis entropy, Nonextensive entropy, Hyperbolicity.\\
%                             
%                             \vfill

%%%%%%%%%%%%%%%%%%%%%%%%%%%%%%%%%%%%%%%%%%%%%%%%%%%%%%%%%%%%%%%%%%

                                                 \centerline{\normalsize\bf INTRODUCTION}

                                                                            \vspace{3mm}

The Havrda-Charvat [1], Dar\'{o}czy [2], Cressie-Read [3], Tsallis [4] (henceforth just ``Tsallis") entropy $S_q [\rho]$  is a single parameter 
$q\in\mathbb{R}$ (called entropic or nonextensive)  family of functionals, which is defined on the space of probability distributions $\rho$ of a 
sample space $\Omega$  (frequently, in Physics, the configuration or phase space of a system) with volume indicated as $vol_{\Omega}$, by
\begin{equation}   
    S_q [\rho]  = k_B \frac{1}{q-1} \left\{1 - \int_{\Omega} [\rho(x)]^q \ dvol_{\Omega} \right\}
\end{equation}
Henceforth the Boltzmann constant will be set $k_B=1$ for simplicity. One can readily see the nonextensive parameter limit
\begin{equation}
     \lim_{q\rightarrow 1}  S_q [\rho] = S_{BGS}  
\end{equation}
where $S_{BGS}$ stands for the well-known Boltzmann/Gibbs/Shannon entropy functional. There is an elaborate thermodynamic formalism 
that has been set up by using $S_q$ that mirrors the one of equilibrium thermodynamics which is based on $S_{BGS}$ [5]. It is somewhat 
unclear, to the author at least,  at the time of this writing what systems are described by the Tsallis entropy. There is a suspicion that 
systems with long-range interactions may be better described by the Tsallis rather than the BGS functional [5], 
but such a conjecture is far from being actually established.  The formalism itself may potentially help clarify the scope and content as well as 
provide a partial answer to this conjecture. To that end it may 
be prudent to explore aspects of $S_q$ by finding the formal differences between $S_q$ and the better known $S_{BGS}$. There is an 
axiomatic formulation of the Tsallis entropy $S_q$ [6], [7] mirroring the one of $S_{BGS}$ that effectively reduces such differences to 
differences in the composition law: for two independent subsystems $\Omega_1, \Omega_2 \subset \Omega$ with marginal probability 
distributions $\rho_1, \rho_2$ the system resulting after their interaction $\Omega_{1\cup 2}$ is described by a probability distribution 
$\rho_{1\cup 2}$ given by the product of the marginals $\rho_{1\cup 2} = \rho_1 \rho_2$. The Tsallis entropy of $\Omega_{1 \cup 2}$ is 
\begin{equation}
 S_q [\rho_{1\cup 2} ] \  = \ S_q [\rho_1] + S_q [\rho_2] + (1-q) S_q [\rho_1] S_q [\rho_2] 
 \end{equation}
As noted above, we believe that understanding the meaning and implications  of (3) may help shed light into what kinds of systems are 
described by $S_q$. \\

                                                                            \vspace{3mm}

%%%%%%%%%%%%%%%%%%%%%%%%%%%%%%%%%%%%%%%%%%%%%%%%%%%%%%%%%%%%%%%%%%%%

                                                  \centerline{\normalsize\bf TSALLIS vs BGS ENTROPY: A METRIC COMPARISON}

                                                                            \vspace{3mm}

Following (3) and in order to ``restore" the additivity of entropies, [8], [9] defined a generalised sum as 
\begin{equation} 
        x \oplus_q y = x + y + (1-q)xy, \hspace{8mm} x, y \in \mathbb{R}    
\end{equation}
Subsequently, a product distributive with respect to $\oplus_q$  was defined in [10], [11]. This allowed [11] to define a ``deformation" of 
$\mathbb{R}$ which was called $\mathbb{R}_q$. Then the differences between $S_{BGS}$ and $S_q$ were recast as differences 
between $\mathbb{R}$ and $\mathbb{R}_q$. We have to point out that the subsequent results assume that $q\in [0, 1)$, even though an 
extension to $q\in\mathbb{R}$ may be possible. To compare $\mathbb{R}$ and $\mathbb{R}_q$  we formed the Cartesian product [12] 
$\mathbb{R} \times \mathbb{R}_q$ and constructed a metric induced by (4) on it. It turned out that this metric is hyperbolic, namely it has 
constant negative (sectional) curvature 
\begin{equation} 
      k = - \{ \log (2-q) \}^2
\end{equation}
Hence the difference between $S_{BGS}$ and $S_q$ is encoded in the metric with curvature (5). By hindsight it is easy to see why (5) arises:
the difference between $S_{BGS}$ and $S_q$ boils down to the difference between the ordinary addition and (4). For values of 
$x, y \in \mathbb{R}$ close to zero, these two additions are essentially the same. If however, both $x$ and $y$ are very large, then 
(4) is essentially a multiplication, rather than an addition. To be more precise, we found [11], [12] the exact form of the isomorphism $\tau_q$ 
between $\mathbb{R}$ and $\mathbb{R}_q$ to be
\begin{equation}
           \tau_q (x) = \frac{(2-q)^x -1}{1-q}
\end{equation} 
which is indeed and exponential map, and its difference from being a linear map becomes far more evident for large $x$. As a result the 
difference between $S_{BGS}$ and $S_q$ should be far more pronounced for highly entropic objects. It may be therefore 
prudent to start seeking the applicability of the Tsallis entropy in systems allowing for such highly entropic distributions. 
The most immediate  example of such a case is a system containing a black hole. Black holes are, by far, the most entropic known 
objects, and even though the statistical origin of this entropy is unclear, at the moment, systems containing them seem to be promising 
grounds for detecting differences between $S_{BGS}$ and $S_q$.   \\   

At this point, we make a rather strong assumption [12]. We  assume henceforth that the hyperbolic metric induced by (4) is the effective metric 
which can be used to analyse the underlying dynamical system. To be more specific, instead of using the expected (Riemannian) metric
in the configuration space of a system, as the metric induced from the kinetic term of the Lagrangian, for instance, we will use its 
``hyperbolization" induced by (4). A possible reason for skepticism to this approach is the following: we logically assume that the underlying 
dynamical system dictates the thermodynamic behaviour of a macroscopic system. This is certainly the viewpoint of Maxwell, Boltzmann etc. 
By doing what we propose, we essentially invert the argument: we let the macroscopic behaviour of a system determine, to some extent, its 
microscopic evolution. This is quite counter-intuitive, and it is unclear to what extent, if any, it is true or not. The advantage of such an 
approach is that the effective behaviour of a system starts manifesting itself even when the number of degrees of freedom is small, therefore it 
may be more accessible to further analysis.\\

As a motivating example consider the ordinary Gaussian distribution. Its use is so widespread due in no small part to the Central Limit 
Theorem.  So, one may start in a system with some probability distribution which in the thermodynamic limit it gives rise to the Gaussian. 
Someone could have reached the same result by starting from the 1-dimensional Gaussian and using its tensorization in the number of 
degrees of freedom. The thermodynamic result would be the same, although details such as the rate of approach to equilibrium may be have 
been different. In the same spirit, we 
pretend that the underlying effective metric of a system is the hyperbolization of the naively chosen Riemannian metric and see 
where it leads us, especially but not necessarily exclusively, in the thermodynamic limit.\\

Following this approach we have been able to demonstrate that if a system is described by the Tsallis entropy and if this is manifest 
through its metric at the configuration / phase space level 
\begin{itemize}
 \item That the largest Lyapunov exponent of the underlying dynamical system should vanish [13],[14]. This can be easily seen: a simple 
          hyperbolic metric keeps distances invariant in a radial direction, according to Gauss' lemma. However it expands them exponentially 
          in transversal directions. For any dynamical system this amounts to shrinking of the geodesic separation at a logarithmic rate with 
          respect to arc-length. Hence any geodesic separation growing exponentially or slower, now becomes polynomial, at best. 
          As a result, the largest Lyapunov exponent which measures the maximum of the geodesic separation in an exponential scale vanishes.  
 \item By a somewhat similar argument which now is applied to volumes instead of distances, we determine that the effective probability      
           distribution that should be used in calculations should not be $\rho$ but its escort $\rho^q$ instead [15]. This does not preclude the two 
           distributions giving the same results in the thermodynamic limit, but when there is discrepancy, the results arising by using the escort 
           should be considered as more appropriate.    
\item  The configuration / phase space volume of a system increases in a power-law rather than the exponential way typical of systems 
           described by the BGS entropy. This result is a straightforward application of the above construction, but it can also be derived by 
           using a sub-Riemannian [16] (3-dimensional Heisenberg group model) of (4) versus the ordinary addition [17].
\item  Using this sub-Riemannian approach one can easily understand Abe's formula for the Tsallis entropy, as a Pansu derivative
           between sub-Riemannian spaces [17].            
\item  One can also easily see the foundations of the Tsallis entropy in the multi-fractal formalism since the underlying geometry is manifestly 
           fractal [17]. This may increase someone's faith in that the assumptions of the previous paragraph, no matter how strong they appear 
            to be, may not be totally unfounded and may lead to reasonable results.            
\item  The Riemannian (hyperbolic) and sub-Riemannian pictures are tied together by examining the hyperbolic nature of the configuration/ 
           phase space with the effective metric and its visual boundary. This allows us to use the very extensive machinery of hyperbolic 
           geometry, not only of Riemannian but more generally of metric spaces [18],[19] in the dynamical analysis of the asymptotic  
           properties of a  system. From this viewpoint, the considerable robustness of the Tsallis entropy already encoded in tis composition (4)  
           can be attributed to the invariance of the induced hyperbolicity of the configuration/phase space under quasi-isometries.   
\item  It is well-known that the Ricci curvature of a Riemannian manifold in some direction controls the rate of change of volumes of  
           small balls locally perpendicular to that direction. Since there is always some ``noise" in the evolution 
           equations and/or the initial conditions of a physical system, the macroscopic behaviour of a system is not actually determined by its 
           configuration (Riemannian) curvature, but actually by its Ricci
           curvature. It is volume deformations of configuration space that matter for the thermodynamic behaviour of a system 
           much more than distance deformations. Therefore all the above comments that apply to induced modifications of the configuration 
           space Riemannian metric of a system due to (4) 
           can now be relaxed somewhat: it is sufficient for someone to demand a deformation of configuration space volumes reflecting (4)
           rather than of actual distances/metrics. When this is implemented, we are naturally lead to a definition of a generalised Ricci curvature 
           called the generalised (Ricci-) Bakry-Emery tensor or $N$-Ricci tensor [20] whose properties are intimately connected to those of the 
           Tsallis entropy (whose properties helped determine it in the first place). We will explore this connection further in an 
           upcoming work [21].                                                    
\end{itemize}

                                                                           \vspace{5mm}

%%%%%%%%%%%%%%%%%%%%%%%%%%%%%%%%%%%%%%%%%%%%%%%%%%%%%%%%%%%%%%%%%%%%                                                                       
                                                                                                                              
                                                        \centerline{\normalsize\bf REFERENCES}
 
                                                                          \vspace{2mm}

 \noindent 1. \ J. Havrda, F. Charvat, \ \emph{Kybernetika} {\bf 3}, \ 30 \ (1967).\\
 \noindent 2. \ Z. Dar\'{o}czy, \ \emph{Inf. Comp. / Inf. Contr.}  {\bf 16}, \ 36 \ (1970).\\
 \noindent 3. \  N.A. Cressie, T. Read, \ \emph{J. Roy. Stat. Soc. B} {\bf 46}, \ 440 \ (1984). \\
 \noindent 4. \  C. Tsallis, \ \emph{J. Stat. Phys.} {\bf 52}, \ 479 \ (1988). \\
 \noindent 5. \  C. Tsallis, \ \emph{Introduction to Nonextensive Statistical Mechanics: Approaching  a Complex 
                           \hspace*{5mm} World}, \ Springer \ (2009).\\ 
 \noindent 6. \  R.J.V. Santos, \ \emph{J. Math. Phys.} {\bf 38}, \ 4104 \ (1997).\\ 
 \noindent 7. \  S. Abe, \ \emph{Phys. Lett. A} {\bf 271}, \ 74 \ (2000).       \\
 \noindent 8. \ L. Nivanen, A. Le Mehaut\'{e}, Q.A. Wang, \ \emph{Rep. Math. Phys.}  {\bf 52}, \ 437 \ (2003).\\ 
 \noindent 9. \ E.P. Borges, \ \emph{Physica A} {\bf 340}, \  95 \  (2004).\\
 \noindent 10. \ T.C. Petit Lob\~{a}o, P.G.S. Cardoso, S.T.R. Pinho, E.P. Borges, \  \emph{Braz. J. Phys.} {\bf 39}, 402 \\
                                  \hspace*{7mm}  (2009). \\          
 \noindent 11. \ N. Kalogeropoulos, \ \emph{Physica A}  {\bf 391}, \ 1120 \ (2012). \\
 \noindent 12. \ N. Kalogeropoulos, \ \emph{Physica A}  {\bf 391}, \ 3435 \ (2012). \\
 \noindent 13. \ N. Kalogeropoulos, \ \emph{QScience Connect} {\bf 12} \ (2012). \\
 \noindent 14. \ N. Kalogeropoulos, \ \emph{Vanishing largest Lyapunov exponent and Tsallis entropy},  {\sf arXiv:1203.2707}\\
 \noindent 15. \ N. Kalogeropoulos, \ \emph{Escort distributions and Tsallis entropy}, \ \  {\sf arXiv:1206.5127}\\ 
 \noindent 16. \ M. Gromov, \  \emph{Carnot-Carath\'{e}odory spaces seen from within}, \ in \ \emph{Sub-Riemannian  \\ 
                               \hspace*{7mm} Geometry,} \  A. Bella\"{i}che, J.-J. Risler (Eds.), \ Birkh\"{a}user \ (1996).\\ 
 \noindent 17. \ N. Kalogeropoulos, \ \emph{Int. J. Geom. Meth. Mod. Phys.} {\bf 10}, \ 1350032 \ (2013).\\
 \noindent 18.  \ M. Gromov, \ \emph{Hyperbolic Groups}, \ in \ \emph{Essays in group theory}, \ S. Gersten (Ed.), \ Math.\\
                              \hspace*{7mm}   Sci. Res. Inst. Publ. {\bf 8}, \ Springer-Verlag \ (1987).\\ 
 \noindent 19.  \ M. Gromov, \emph{Asymptotic invariants of infinite groups},  in  \emph{Geometric group theory, Vol 2},\\ 
                                \hspace*{7mm}  G.A. Niblo, M.A. Roller (Eds.), Cambridge University Press (1993).\\ 
 \noindent 20. \ J. Lott, C. Villani, \ \emph{Ann. Math.} {\bf 169}, \ 903 \ (2009).\\
 \noindent 21. \  N. Kalogeropoulos,\ \  \emph{In preparation}.\\
\end{document}